\documentclass[aps,prc,preprintnumbers,superscriptaddress,showpacs,floatfix]{revtex4}
\usepackage{amsmath}
\usepackage{epsfig}
\usepackage{bm}
\usepackage{amssymb}

\begin{document}

\title{On the fluid behavior of a baryon rich hadron resonance gas}
\author{Gabriel~S.~Denicol}
\author{Charles Gale}
\author{Sangyong Jeon}
\affiliation{Department of Physics, McGill University, 3600 University Street, Montreal,
Quebec, H3A 2T8, Canada}
\author{Jorge Noronha}
\affiliation{Instituto de F\'{\i}sica, Universidade de S\~{a}o Paulo, C.P. 66318,
05315-970 S\~{a}o Paulo, SP, Brazil}

\begin{abstract}
We investigate the effects of finite baryon chemical potential on the
transport properties of a hadron resonance gas. We find that a hadron
resonance gas with large baryon number density is closer to the ideal fluid
limit than the corresponding gas with zero baryon number. This suggests that
the system created at the Relativistic Heavy Ion Collider (RHIC) at lower
collision energies may behave as a fluid, with an effective fluidity close to the one found at RHIC's highest energy near phase transition. This might explain
why the differential elliptic flow coefficient measured at lower collisional
energies at RHIC is similar to the one observed at high energies.
\end{abstract}

\pacs{12.38.Mh, 24.10.Pa, 24.85.+p, 25.75.Dw}
\maketitle


\section{Introduction}

The main goal of ultrarelativistic heavy ion collisions is to create and
study new states of nuclear matter. The Relativistic Heavy Ion Collider
(RHIC) and the Large Hadron Collider (LHC) are able to reach collisional
energies high enough to create the Quark-Gluon Plasma (QGP) \cite{review1},
a new state of nuclear matter in which the quarks and gluons are the
relevant degrees of freedom \cite{whitepapers}. One of the most surprising
discoveries made at RHIC and later confirmed at the LHC is that the QGP
seems to behave as a nearly ideal fluid, with one of the smallest values for
the shear viscosity to entropy density ratio, $\eta /s$, ever observed in
nature \cite{review2}.

In the last two years, RHIC started to investigate collisions at lower
energies, the so-called RHIC low energy scan, in order to probe the phase
diagram of nuclear matter at large baryon chemical potentials and to study
the existence of a critical point \cite{Stephanov:1998dy}. A striking
discovery made by the low energy scan program was that the differential
elliptic flow of charged hadrons, $v_{2}(p_{T})$, does not change
considerably as one goes to lower collisional energy \cite{v2_lowenergy}.
This came as a surprise since the large values of $v_{2}(p_{T})$ observed at
RHIC's highest energy were taken as the main evidence for the ``almost"
ideal fluid nature of the QGP \cite{review2}. The fact that such a high
differential flow coefficient is also observed at smaller collisional
energies, in which the QGP is only present during a very small fraction of
the total lifetime of the system, may suggest that the hadron gas formed at
the end of the collision can also exhibit strong collective flow. Several
calculations \cite%
{Shen:2012vn,Auvinen:2013qia,Plumari:2013bga,Solanki:2012ne} based on
fluid-dynamical and/or transport models performed at lower RHIC energies
have tried to understand how such large differential elliptic flow can occur
even at these energies. However, at nonzero baryon chemical potential, the
uncertainties in the equation of state and in the transport coefficients,
and the existence of baryon stopping in the initial stages render such a
task more complicated.

Recently, it was shown that the dissipative effects on $v_{2}(p_{T})$ at
RHIC's top energy originate mostly from the viscosity of the QGP around the
(pseudo) phase transition and the viscosity of the hadron gas formed
afterwards \cite{niemi_PRL}. As one goes to higher energies, $v_{2}(p_{T})$
becomes more sensitive to the viscosity of the QGP phase (which is expected
to increase with temperature), while if one goes to smaller energies, this
observable becomes more and more sensitive to the viscosity of the hadron
gas (which is expected to increase with decreasing temperature). Since $%
\eta/s$ increases with decreasing temperature in the hadron gas phase, one
would naively expect $v_{2}(p_{T})$ to become considerably smaller as the
initial collisional energy of the system is reduced.

Note that the aforementioned behavior of $\eta /s$ as a function of energy
was only estimated for nuclear systems at zero baryon chemical potential 
\cite{Prakash,Noronha,Wiranata}. Therefore, it is important to understand
how this is modified in the presence of a finite baryon chemical potential, $%
\mu _{B}$. Previous works already indicate that the $\eta /s$ of a gas is
reduced at finite chemical potential \cite{Bass_UrQMD, Japonese}. For
example, $\eta /s$ was computed for a pion-nucleon gas in Ref.~\cite%
{Japonese} using the Chapman-Enskog formalism \cite{CE} and it was shown to
be reduced when $\mu _{B}>0$. The exception is in Ref.~\cite{Hauer}, where
the $\eta /s$ of a hadron resonance gas was estimated in the excluded volume
hadron gas model using a simplified \textit{Ansatz} for the shear viscosity
coefficient. In that work the result obtained was that $\eta /s$ is
increased when the baryon chemical potential is nonzero.

On the other hand, it is important to state here that the fluidity of a
system at finite chemical potential cannot be inferred by the smallness of $%
\eta /s$, as was argued in Ref.\ \cite{koch}. A more realistic measure of
fluidity would be the ratio of shear viscosity over the enthalpy multiplied
by the temperature, $\eta T/(\epsilon +p)$.\ Note that, when the chemical
potential is zero, this quantity reduces to the ratio between shear
viscosity and entropy density. On the other hand, when $\mu _{B}$ is large $%
\eta T/(\epsilon +p)$ considerably differs from $\eta /s$, which can make a
difference when analyzing how well a certain system flows (the same can be
said for systems that are chemically frozen).

In this work, we compute the shear viscosity of a hadron resonance gas at
nonzero temperature and baryon chemical potential using Chapman-Enskog
theory \cite{CE,prd} within a simple hadronic model \cite{DeGroot}. We then
compute $\eta T/(\epsilon +p)$ for this system and show that it is
considerably reduced when the baryon chemical potential is nonzero. For the
values of baryon chemical potential expected to occur around the (chemical
freeze-out) phase transition of QCD \cite%
{Andronic:2008gu,Andronic:2005yp,Cleymans:1998fq,Cleymans:1999st,Cleymans:2005xv,NoronhaHostler:2009tz}%
, this reduction is shown to be particularly large. This result suggests
that the baryon rich hadron gas formed in the intermediate energies obtained
at RHIC, and at future facilities such as FAIR \cite{FAIR}, can display
strong collective flow, as already implied by the flow harmonic data
obtained at RHIC.

In this work the metric tensor is $g^{\mu \nu }\equiv \mathrm{diag}(+,-,-,-)$
and natural units are employed, $\hbar =c=k_{B}=1$. Furthermore, we limit
our discussion to Boltzmann gases.

\section{Formalism}

We start by generalizing the method proposed in Ref.~\cite{prd} to calculate
the retarded Green's function associated with the shear-stress tensor for
the case of a dilute multi-component system. As shown in Ref.~\cite{prd},
this formalism will give the same answer as the Chapman-Enskog expansion 
\cite{CE} for the linear coefficients of the Burnett theory \cite{Burnett},
i.e., the gradient expansion.

We perform our calculations in the local rest frame of the fluid where $%
u^{\mu }=(1,0,0,0)$. Since we restrict our discussion to shear viscosity, we
may consider the simplified scenario in which the fluid does not accelerate, 
$u^{\mu }\partial _{\mu }u^{\nu }\equiv 0$, nor expands, $\partial _{\mu
}u^{\mu }\equiv 0$, and the temperature, $T=\beta _{0}^{-1}$, and chemical
potential, $\mu ^{i}=\alpha _{0}^{i}/\beta _{0}$, do not vary in space and
time. Here, $\mu ^{i}=b_{i}\mu _{B}$, where $\mu _{B}$ is the baryon
chemical potential and $b_{i}$ is the baryon number of the $i$--th hadron
species. For the sake of simplicity, we only consider classical statistics.

After these simplifications, the Boltzmann equation linearized around a
classical equilibrium state, $f_{0\mathbf{p}}^{\left( i\right) }=\exp
(\alpha _{0}^{i}-\beta _{0}E_{\mathbf{k}i})$, becomes \cite{prd}%
\begin{equation}
\partial _{t}\delta f_{\mathbf{k}}^{\left( i\right) }+\mathbf{v}_{i}\cdot
\nabla \delta f_{\mathbf{k}}^{\left( i\right) }-\sum\limits_{j=1}^{N_{%
\mathrm{spec}}}\hat{C}_{ij}\left( K_{i}\right) \delta f_{\mathbf{k}}^{\left(
i\right) }=\beta _{0}E_{\mathbf{k}i}^{-1}f_{0\mathbf{p}}^{\left( i\right)
}p_{i}^{\left\langle \mu \right. }p_{i}^{\left. \nu \right\rangle }\sigma
_{\mu \nu },  \label{linearboltzmann}
\end{equation}%
where $E_{\mathbf{k}i}$ is the energy of the particle, $N_{\mathrm{spec}}$
is the number of hadron species considered, $\nabla _{\mu }=\Delta _{\mu
}^{\nu }\partial _{\nu }$ is the spatial derivative, and $\sigma _{\mu \nu
}=\nabla _{\left\langle \mu \right. }u_{\left. \nu \right\rangle }$ is the
shear tensor. We defined $\Delta _{\nu }^{\mu }\equiv g_{\nu }^{\mu }-u^{\mu
}u_{\nu }$, $\mathbf{v}_{i}\equiv \mathbf{k/}E_{\mathbf{k}i}$, $\delta f_{%
\mathbf{k}}^{\left( i\right) }\equiv f_{\mathbf{k}}^{\left( i\right) }-f_{0%
\mathbf{k}}^{\left( i\right) }$, and $A^{\left\langle \mu \nu \right\rangle
}=\Delta ^{\mu \nu \alpha \beta }A_{\alpha \beta }$ where $\Delta ^{\mu \nu
\alpha \beta }=\left( \Delta ^{\mu \alpha }\Delta ^{\nu \beta }+\Delta ^{\mu
\beta }\Delta ^{\nu \alpha }\right) /2-\Delta ^{\mu \nu }\Delta ^{\alpha
\beta }/3$ is the doubly symmetric, traceless projection operator. Also, we
introduced the linearized collision operator, $\hat{C}_{ij}$,%
\begin{equation}
\hat{C}_{ij}\delta f_{\mathbf{k}}^{\left( i\right) }=\int dK_{j}^{\prime
}dP_{i}dP_{j}^{\prime }\gamma _{ij}W_{\mathbf{kk}^{\prime }-\mathbf{pp}%
^{\prime }}^{ij}E_{\mathbf{p}i}^{-1}f_{0\mathbf{k}}^{\left( i\right) }f_{0%
\mathbf{k}^{\prime }}^{\left( j\right) }\left( \frac{\delta f_{\mathbf{p}%
}^{\left( i\right) }}{f_{0\mathbf{p}}^{\left( i\right) }}+\frac{\delta f_{%
\mathbf{p}^{\prime }}^{\left( j\right) }}{f_{0\mathbf{p}^{\prime }}^{\left(
j\right) }}-\frac{\delta f_{\mathbf{k}}^{\left( i\right) }}{f_{0\mathbf{k}%
}^{\left( i\right) }}-\frac{\delta f_{\mathbf{k}^{\prime }}^{\left( j\right)
}}{f_{0\mathbf{k}^{\prime }}^{\left( j\right) }}\right) .
\label{collisionterm}
\end{equation}%
Here, $dK_{i}\equiv g_{i}d^{3}\vec{k}/\left[ (2\pi )^{3}k_{i}^{0}\right] $
is the Lorentz-invariant measure, $g_{i}$ is the degeneracy factor of the $i$%
--th hadron species, $\gamma _{ij}=1-\left( 1/2\right) \delta _{ij}$, and $%
g_{i}g_{j}W_{\mathbf{kk}\prime \rightarrow \mathbf{pp}\prime }^{ij}=s\sigma
_{ij}\left( 2\pi \right) ^{6}\delta ^{4}\left( p_{i}+p_{j}^{\prime
}-k_{i}-k_{j}^{\prime }\right) $ is the corresponding transition rate, with $%
\sigma _{ij}$ being the differential cross section and $s$ the Mandelstan
variable.

Using Eqs.\ (\ref{linearboltzmann}) and (\ref{collisionterm}), we express
the Fourier transform of the shear-stress tensor, $\tilde{\pi}^{\mu \nu }(Q)$%
, in terms of the Fourier transform of the shear tensor, $\tilde{\sigma}%
_{\alpha \beta }(Q)$, in the traditional form of linear response $\tilde{\pi}%
^{\mu \nu }=G_{R}^{\mu \nu \alpha \beta }\tilde{\sigma}_{\alpha \beta }$ 
\cite{prd}. The Green's function has the following structure%
\begin{equation}
\tilde{G}_{R}^{\mu \nu \alpha \beta }\left( Q\right) =\sum\limits_{i=1}^{N_{%
\mathrm{spec}}}\int dK_{i}\,k_{i}^{\left\langle \mu \right. }k_{i}^{\left.
\nu \right\rangle }\frac{\beta _{0}}{-i\omega +i\mathbf{v}_{i}\cdot \mathbf{q%
}-\sum\limits_{j=1}^{N_{\mathrm{spec}}}\hat{C}_{ij}}f_{0\mathbf{k}}^{\left(
i\right) }E_{i\mathbf{k}}^{-1}k_{i}^{\left\langle \alpha \right.
}k_{i}^{\left. \beta \right\rangle }\,,  \label{oh_man}
\end{equation}%
where $Q^{\mu }=(\omega ,\mathbf{q})$ and we used the definition of the
total shear-stress tensor of a multi-component system%
\begin{equation}
\pi ^{\mu \nu }=\sum\limits_{i=1}^{N_{\mathrm{spec}}}\int
dK_{i}\,k_{i}^{\left\langle \mu \right. }k_{i}^{\left. \nu \right\rangle
}\delta f_{\mathbf{k}}^{\left( i\right) }.  \label{shear}
\end{equation}

Our goal is to compute the above retarded Green's function and its
derivatives at vanishing wavenumber. This will be enough to determine the
relevant transport coefficients of linearized relativistic fluid dynamics.
Following the strategy outlined in \cite{prd}, we start by introducing the
irreducible second rank tensor $B^{\alpha \beta }\left( Q,K_{i}\right) $,
which satisfies the following integro-differential equation,%
\begin{equation}
\left( -i\omega +i\mathbf{v}_{i}\cdot \mathbf{q}-\sum\limits_{j=1}^{N_{%
\mathrm{spec}}}\hat{C}_{ij}\right) B^{\alpha \beta }\left( Q,K_{i}\right)
=\beta _{0}E_{i\mathbf{k}}^{-1}k_{i}^{\left\langle \alpha \right.
}k_{i}^{\left. \beta \right\rangle }f_{0\mathbf{k}}^{\left( i\right) }.
\label{equation}
\end{equation}%
In general, $B^{\alpha \beta }$ is a function of $Q$ but, as mentioned
above, we shall only need it at vanishing wavenumber, $\mathbf{q}=0$. Then $%
B^{\alpha \beta }=B^{\alpha \beta }\left( \omega ,K_{i}\right) $ and its
dependence on $k_{i}^{\mu }$ can be described by the following expansion, 
\begin{equation}
B^{\alpha \beta }\left( \omega ,K\right) =f_{0\mathbf{k}}^{\left( i\right)
}k_{i}^{\left\langle \alpha \right. }k_{i}^{\left. \beta \right\rangle
}\sum_{n=0}^{\infty }a_{n}^{\left( i\right) }\left( \omega \right) E_{i%
\mathbf{k}}^{n}.  \label{expansion}
\end{equation}%
Substituting Eq.\ (\ref{expansion}) into Eq.\ (\ref{oh_man}), it follows
that 
\begin{eqnarray}
\tilde{G}_{R}^{\mu \nu \alpha \beta }\left( \omega ,\mathbf{0}\right)
&=&\sum\limits_{i=1}^{N_{\mathrm{spec}}}\sum_{n=0}^{\infty }a_{n}^{\left(
i\right) }\left( \omega \right) \int dK_{i}\,k_{i}^{\left\langle \mu \right.
}k_{i}^{\left. \nu \right\rangle }k_{i}^{\left\langle \alpha \right.
}k_{i}^{\left. \beta \right\rangle }E_{i\mathbf{k}}^{n}f_{0\mathbf{k}%
}^{\left( i\right) }  \notag \\
&=&2\Delta ^{\mu \nu \alpha \beta }\sum\limits_{i=1}^{N_{\mathrm{spec}%
}}\sum_{n=0}^{\infty }I_{n+4,2}^{\left( i\right) }\;a_{n}^{\left( i\right)
}\left( \omega \right) \,,
\end{eqnarray}%
where we used \cite{DeGroot} 
\begin{equation}
\int dK_{i}\,k_{i}^{\left\langle \mu \right. }k_{i}^{\left. \nu
\right\rangle }k_{i}^{\left\langle \alpha \right. }k_{i}^{\left. \beta
\right\rangle }E_{i\mathbf{k}}^{n}f_{0\mathbf{k}}^{\left( i\right) }=\frac{2%
}{5!!}\,\Delta ^{\mu \nu \alpha \beta }\int dK\,_{i}E_{i\mathbf{k}}^{n}f_{0%
\mathbf{k}}^{\left( i\right) }\left( m_{i}^{2}-E_{i\mathbf{k}}^{2}\right)
^{2}\;.  \label{56}
\end{equation}%
Then, the linear relation between $\pi ^{\mu \nu }$ and $\sigma ^{\mu \nu }$
can be cast into a more convenient form, $\tilde{\pi}^{\mu \nu }\left(
\omega ,\mathbf{0}\right) =2\tilde{G}_{R}\left( \omega ,\mathbf{0}\right) 
\tilde{\sigma}^{\mu \nu }\left( \omega ,\mathbf{0}\right) $, where we
introduced the retarded Green's function%
\begin{equation}
\tilde{G}_{R}\left( \omega ,\mathbf{0}\right) =\sum\limits_{i=1}^{N_{\mathrm{%
spec}}}\sum_{n=0}^{\infty }I_{n+4,2}^{\left( i\right) }\;a_{n}^{\left(
i\right) }\left( \omega \right) \text{.}
\end{equation}

Equations such as (\ref{equation}) appear quite often in problems involving
the extraction of transport coefficients from the Boltzmann equation. The
way to solve this equation is to substitute the expansion (\ref{expansion})
into Eq.\ (\ref{equation}), multiply by the basis component chosen to expand 
$B^{\alpha \beta }$, i.e., multiply by $E_{i\mathbf{k}}^{m}k_{i}^{\left%
\langle \mu \right. }k_{i}^{\left. \nu \right\rangle }$, and integrate over $%
dK_{i}$. Then one obtains 
\begin{eqnarray}
&&\sum\limits_{i=1}^{N_{\mathrm{spec}}}\sum_{n=0}^{\infty }\int dK_{i}E_{i%
\mathbf{k}}^{m}k_{i}^{\left\langle \mu \right. }k_{i}^{\left. \nu
\right\rangle }\left[ -i\omega -\sum\limits_{j=1}^{N_{\mathrm{spec}}}\hat{C}%
_{ij}\right] a_{n}^{\left( i\right) }\left( \omega \right) f_{0\mathbf{k}%
}^{\left( i\right) }k_{i}^{\left\langle \alpha \right. }k_{i}^{\left. \beta
\right\rangle }E_{i\mathbf{k}}^{n}  \notag \\
&=&\beta _{0}\sum\limits_{i=1}^{N_{\mathrm{spec}}}\int dK_{i}E_{i\mathbf{k}%
}^{m}k_{i}^{\left\langle \mu \right. }k_{i}^{\left. \nu \right\rangle }E_{i%
\mathbf{k}}^{-1}k_{i}^{\left\langle \alpha \right. }k_{i}^{\left. \beta
\right\rangle }f_{0\mathbf{k}}^{\left( i\right) }.
\end{eqnarray}%
The function $a_{n}\left( \omega \right) $ can be solved by inverting the
above equation.

For any practical calculation, however, only a finite number of basis
elements are used in the expansion of $B^{\alpha \beta }$. Truncating such
momentum expansion is a common approach in kinetic theory and happens, for
example, when dealing with the Chapmann-Enskog expansion \cite{CE}. Here we
use the simplest truncation possible in which only one member of the basis
is used (this is equivalent to the 14-moment approximation in the method of
moments \cite{prd,IS,DKR}). This truncation should still provide transport
coefficients that are accurate up to $\sim 10\%$ (see, for instance, Refs.\ 
\cite{DeGroot,DNMR}). In such simplified approximation,%
\begin{equation}
\sum\limits_{j=1}^{N_{\mathrm{spec}}}\left[ -i\omega I_{42}^{\left( i\right)
}\delta ^{ij}+\mathcal{A}^{\left( i\right) }\delta ^{ij}+\mathcal{C}^{\left(
ij\right) }\right] a_{0}^{\left( j\right) }\left( \omega \right) =\beta
_{0}I_{32}^{\left( i\right) },
\end{equation}%
where%
\begin{eqnarray}
\mathcal{A}^{\left( i\right) } &=&-\sum\limits_{j=1}^{N_{\mathrm{spec}}}%
\frac{1}{10}\int dK_{i}dK_{j}^{\prime }dP_{i}dP_{j}^{\prime }\gamma _{ij}W_{%
\mathbf{kk}^{\prime }-\mathbf{pp}^{\prime }}^{ij}E_{i\mathbf{k}%
}^{-1}k_{i}^{\left\langle \mu \right. }k_{i}^{\left. \nu \right\rangle }f_{0%
\mathbf{p}}^{\left( i\right) }f_{0\mathbf{p}^{\prime }}^{\left( j\right)
}\left( k_{i\left\langle \mu \right. }k_{i\left. \nu \right\rangle
}-p_{i\left\langle \mu \right. }p_{i\left. \nu \right\rangle }\right) ,
\label{int1} \\
\mathcal{C}^{\left( ij\right) } &=&-\frac{1}{10}\int dK_{i}dK_{j}^{\prime
}dP_{i}dP_{j}^{\prime }\gamma _{ij}W_{\mathbf{kk}^{\prime }-\mathbf{pp}%
^{\prime }}^{ij}E_{i\mathbf{k}}^{-1}k_{i}^{\left\langle \mu \right.
}k_{i}^{\left. \nu \right\rangle }f_{0\mathbf{p}}^{\left( i\right) }f_{0%
\mathbf{p}^{\prime }}^{\left( j\right) }\left( k_{j\left\langle \mu \right.
}^{\prime }k_{j\left. \nu \right\rangle }^{\prime }-p_{j\left\langle \mu
\right. }^{\prime }p_{j\left. \nu \right\rangle }^{\prime }\right) .
\label{int2}
\end{eqnarray}%
The formal solution for $a_{0}^{\left( i\right) }\left( \omega \right) $ is 
\begin{equation}
a_{0}^{\left( i\right) }\left( \omega \right) =\beta
_{0}\sum\limits_{j=1}^{N_{\mathrm{spec}}}\left[ \left( -i\omega
I_{42}^{\left( i\right) }\mathbf{\hat{1}}+\mathcal{A}^{\left( i\right) }%
\mathbf{\hat{1}}+\mathbf{\hat{C}}\right) ^{-1}\right] ^{ij}I_{32}^{\left(
j\right) },
\end{equation}%
where $\mathbf{\hat{C}}^{ij}=\mathcal{C}^{\left( ij\right) }$, and the
retarded Green's function, $\tilde{G}_{R}\left( \omega ,\mathbf{0}\right) $,
becomes%
\begin{equation}
\tilde{G}_{R}\left( \omega ,\mathbf{0}\right) =\beta
_{0}\sum\limits_{i=1}^{N_{\mathrm{spec}}}\sum\limits_{j=1}^{N_{\mathrm{spec}%
}}I_{42}^{\left( i\right) }\left[ \left( -i\omega I_{42}^{\left( i\right) }%
\mathbf{\hat{1}}+\mathcal{A}^{\left( i\right) }\mathbf{\hat{1}}+\mathbf{\hat{%
C}}\right) ^{-1}\right] ^{ij}I_{32}^{\left( j\right) }.
\end{equation}%
The derivative of this Green's function at zero frequency is%
\begin{eqnarray*}
\left. \partial _{\omega }\tilde{G}_{R}\left( \omega ,\mathbf{0}\right)
\right\vert _{\omega =0} &=&\sum\limits_{i=1}^{N_{\mathrm{spec}%
}}I_{42}^{\left( i\right) }\;\left. \partial _{\omega }a_{0}^{\left(
i\right) }\left( \omega \right) \right\vert _{\omega =0}\text{,} \\
&=&i\beta _{0}\sum\limits_{i,j,m=1}^{N_{\mathrm{spec}}}I_{42}^{\left(
i\right) }\;\left[ \left( \mathcal{A}^{\left( i\right) }\mathbf{\hat{1}}+%
\mathbf{\hat{C}}\right) ^{-1}\right] ^{ij}I_{42}^{\left( j\right) }\left[
\left( \mathcal{A}^{\left( i\right) }\mathbf{\hat{1}}+\mathbf{\hat{C}}%
\right) ^{-1}\right] ^{jm}I_{32}^{\left( m\right) }\text{.}
\end{eqnarray*}

The linearized Burnett equation emerging from the above Green's function is%
\begin{equation*}
\pi ^{\mu \nu }=2\eta \sigma ^{\mu \nu }-2\lambda \Delta _{\alpha \beta
}^{\mu \nu }u^{\lambda }\partial _{\lambda }\sigma ^{\alpha \beta }+\left( %
\mbox{terms of higher order}\right) ,
\end{equation*}%
where the shear viscosity coefficient is identified as 
\begin{equation}
\eta \equiv \left. \tilde{G}_{R}\left( \omega ,\mathbf{0}\right) \right\vert
_{\omega =0}=\beta _{0}\sum\limits_{i=1}^{N_{\mathrm{spec}%
}}\sum\limits_{j=1}^{N_{\mathrm{spec}}}I_{42}^{\left( i\right) }\left[
\left( \mathcal{A}^{\left( i\right) }\mathbf{\hat{1}}+\mathbf{\hat{C}}%
\right) ^{-1}\right] ^{ij}I_{32}^{\left( j\right) }\,,  \label{eta}
\end{equation}%
and the transport coefficient of the second order term $2\Delta _{\alpha
\beta }^{\mu \nu }u^{\lambda }\partial _{\lambda }\sigma ^{\alpha \beta }$,
here referred to as $\lambda $, is given by%
\begin{equation}
\lambda =-i\left. \partial _{\omega }\tilde{G}_{R}\left( \omega ,\mathbf{0}%
\right) \right\vert _{\omega =0}.  \label{lambda}
\end{equation}%
Nonlinear terms that may be included in the gradient expansion are not
considered in this work.

\section{Thermodynamics}

In the following, we briefly discuss how thermodynamic quantities are
computed in a multi-component gas. The baryon number density, $n_{B}$, the
energy density, $\varepsilon $, and the thermodynamic pressure, $P$, are
computed using the hadron resonance gas model,%
\begin{eqnarray}
n_{B} &=&\sum_{i=1}^{N_{\mathrm{spec}}}b_{i}\int dK_{i}E_{i\mathbf{k}}\exp
\left( -\beta _{0}\,E_{i\mathbf{k}}+\alpha _{0}^{i}\right) \text{ }\,, 
\notag \\
\varepsilon &=&\sum_{i=1}^{N_{\mathrm{spec}}}\int dK_{i}E_{i\mathbf{k}%
}^{2}\exp \left( -\beta _{0}\,E_{i\mathbf{k}}+\alpha _{0}^{i}\right) \text{ }%
,  \notag \\
P &=&\frac{1}{3}\sum_{i=1}^{N_{\mathrm{spec}}}\int dK_{i}\left( E_{i\mathbf{k%
}}^{2}-m_{i}^{2}\right) \exp \left( -\beta _{0}\,E_{i\mathbf{k}}+\alpha
_{0}^{i}\right) \;.  \label{def_hy_qua}
\end{eqnarray}%
The entropy density, $s$, is computed using the following thermodynamic
relation%
\begin{equation*}
\varepsilon +P=Ts+\mu _{B}n_{B}\text{ }.
\end{equation*}%
The enthalpy, $h$, is defined as%
\begin{equation*}
h=\varepsilon +P\text{ }.
\end{equation*}%
Note that the ratio of enthalpy to temperature is equivalent to the entropy
density when the baryon chemical potential is zero, i.e., when $\mu _{B}=0$, 
$h=Ts$ .

\begin{figure}[tbp]
\begin{minipage}{.45\linewidth}
\hspace{-1.5cm}
\includegraphics[width=1.1\textwidth]{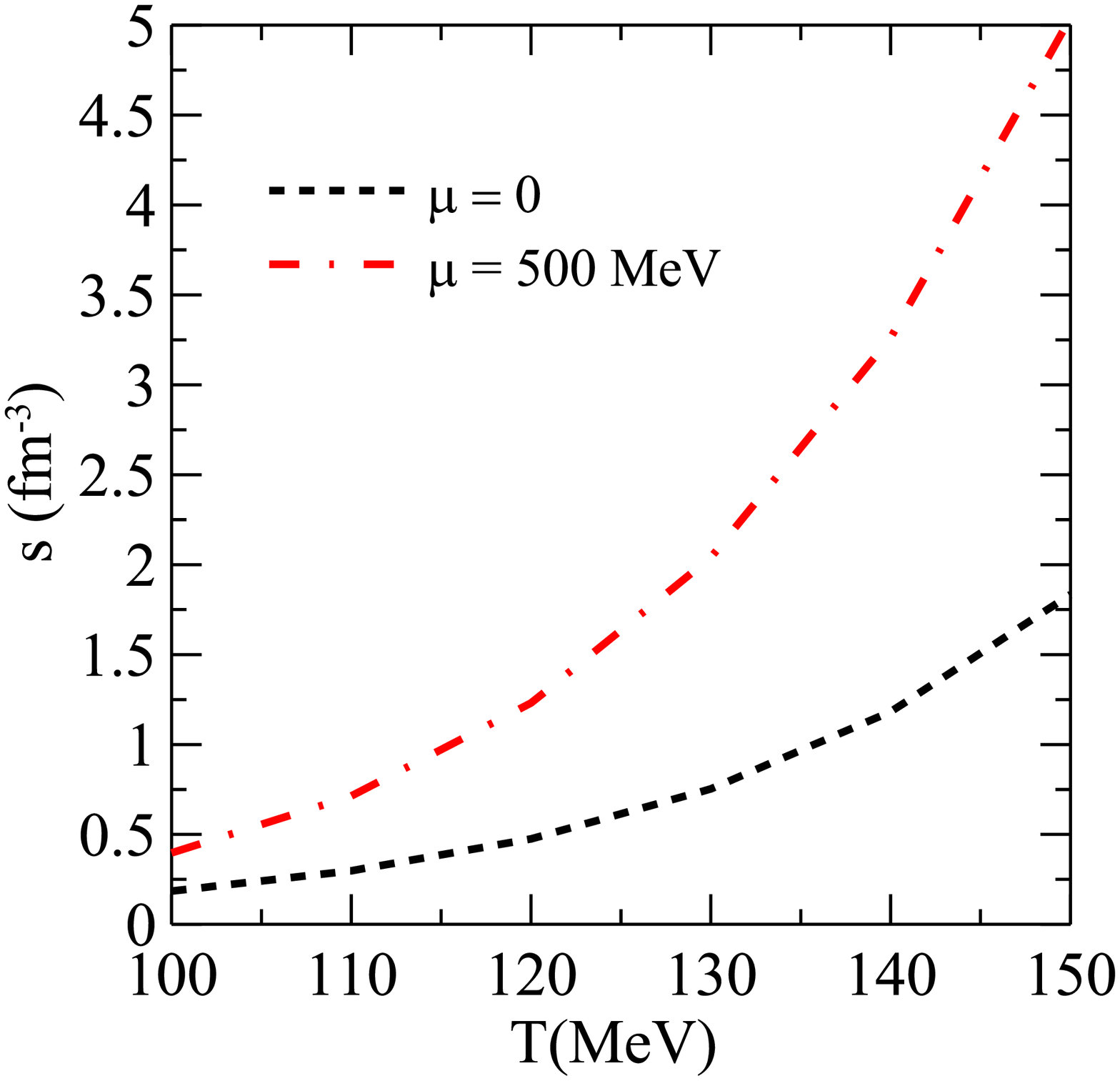} 
\end{minipage}  
\begin{minipage}{.45\linewidth}
\hspace{-1.5cm}
\includegraphics[width=1.1\textwidth]{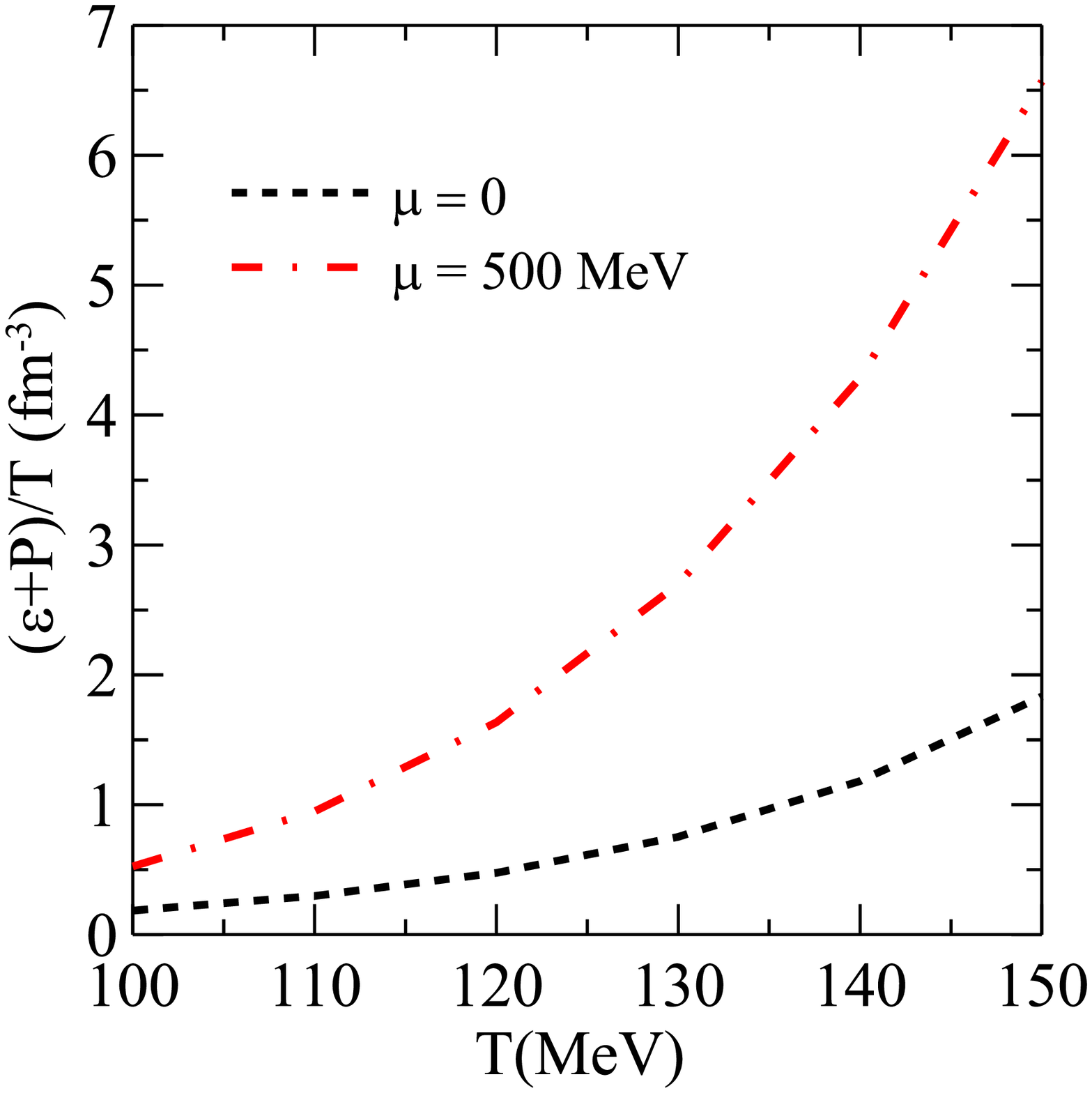} 
\end{minipage} 
\caption{The entropy density, $s$, (left panel) and the ratio of enthalpy to
temperature $h/T$ (right panel) as a function of temperature (in MeV) for $%
\protect\mu _{B}=0$ (black dotted line) and $\protect\mu _{B}=500$ MeV (red
dash-dotted line).}
\label{fig0}
\end{figure}

In Fig.\ \ref{fig0} we show the entropy density and the ratio of enthalpy to
temperature as a function of temperature for $\mu _{B}=0$ and $\mu _{B}=0.5$
GeV. It is clear that both these thermodynamic quantities increase in the
presence of a positive baryon chemical potential, with the increase in $h/T$
being more pronounced.

\section{Results}

The purpose of this work is not to compute the transport properties of a
hadron gas with precision. In order to accomplish this, one would have to
know the cross sections of all processes involving the known hadrons and
resonances as well as include the effect of Hagedorn states and repulsive
interactions. Currently, the inclusion of all these effects with precision
is just not possible. What we actually aim to investigate is how the
fluidity of such a gas is modified when the baryon chemical potential is
present. This effect can be consistently estimated with a more simple model
of a hadron gas in which all cross sections are constant and fixed to a
certain value. Here, we shall consider two cases: (i) assume that all
hadrons have a $1$ fm radius, leading to total cross-sections of $\sigma
_{T}=\pi $ fm$^{2}$ and (ii) assume that meson-meson, meson-baryon, and
baryon-baryon cross sections scale as $4:6:9$, respectively, with the
baryon-baryon cross section beying set to $\pi $ fm$^{2}$. That is, in case
(ii) $\sigma _{\mathrm{meson-meson}}=4/9\times \sigma _{\mathrm{baryon-baryon%
}}$ and $\sigma _{\mathrm{meson-baryon}}=2/3\times \sigma _{\mathrm{%
baryon-baryon}}$. Such simplified systems can be used to estimate the
effects of a nonzero baryon chemical potential on the transport coefficients
of a hadron resonance gas. For constant cross-sections, it is
straightforward to solve integrals (\ref{int1}) and (\ref{int2}). Then, $%
\eta $ and $\lambda $ can be computed from Eqs.\ (\ref{eta}) and (\ref%
{lambda}), while all thermodynamic quantities are computed from the formulas
specified in the previous section.

In the following, we shall consider $3$ different choices of baryon chemical
potential: $\mu _{B}=0$, $\mu _{B}=500$ MeV, and the $\mu _{B}(T)$ obtained
from thermal fits to heavy ion collisions at several collisional energies.
In Ref. \cite{Cleymans:2005xv}, the following parametrization of $T\left(
\mu _{B}\right) $ was extracted from thermal fit calculations,%
\begin{equation}
T\left( \mu _{B}\right) \approx a-b\mu _{B}^{2}-c\mu _{B}^{4},  \label{T_mu}
\end{equation}%
where $a\simeq 0.166$ GeV, $b\simeq 0.139$ GeV$^{-1}$, and $c\simeq 0.053$
GeV$^{-3}$. This parametrization provides the temperature as a function of
baryon chemical potential in the chemical freeze-out transition. In
practice, the chemical freeze-out transition is very close to the actual
(pseudo) phase transition region and we will use this parametrization to
estimate $T\left( \mu _{B}\right) $ near the phase transition region. Here,
we just invert Eq. (\ref{T_mu}) to obtain $\mu _{B}\left( T\right) $.
Finally, we consider all hadrons and resonances with masses up to $2$ GeV,
where the masses and degeneracy factors of each hadron/resonance are taken
from Ref.\ \cite{pdg}.

\begin{figure}[tbp]
\begin{minipage}{.45\linewidth}
\hspace{-1.5cm}
\includegraphics[width=1.1\textwidth]{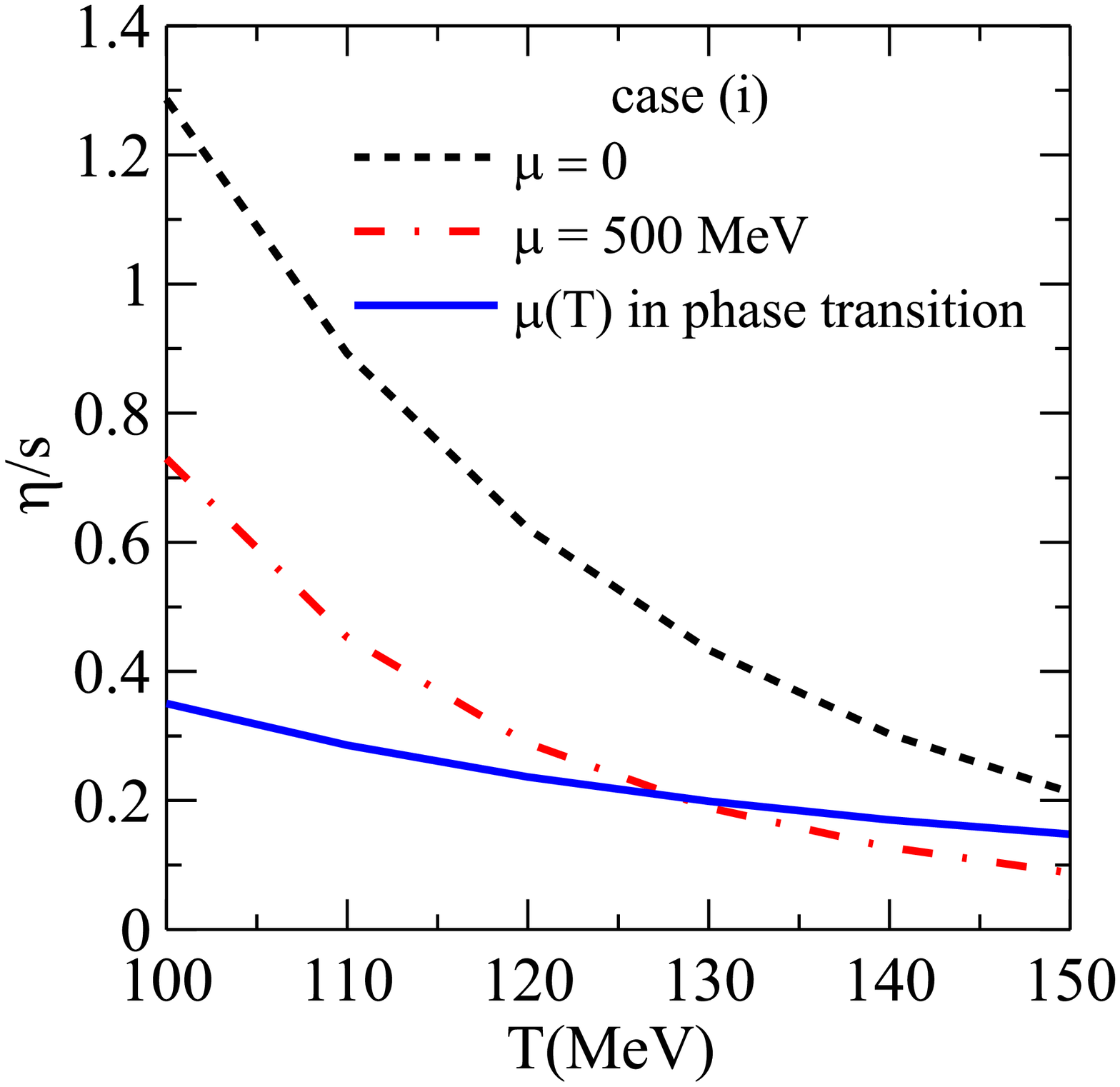} 
\end{minipage}  
\begin{minipage}{.45\linewidth}
\hspace{-1.5cm}
\includegraphics[width=1.1\textwidth]{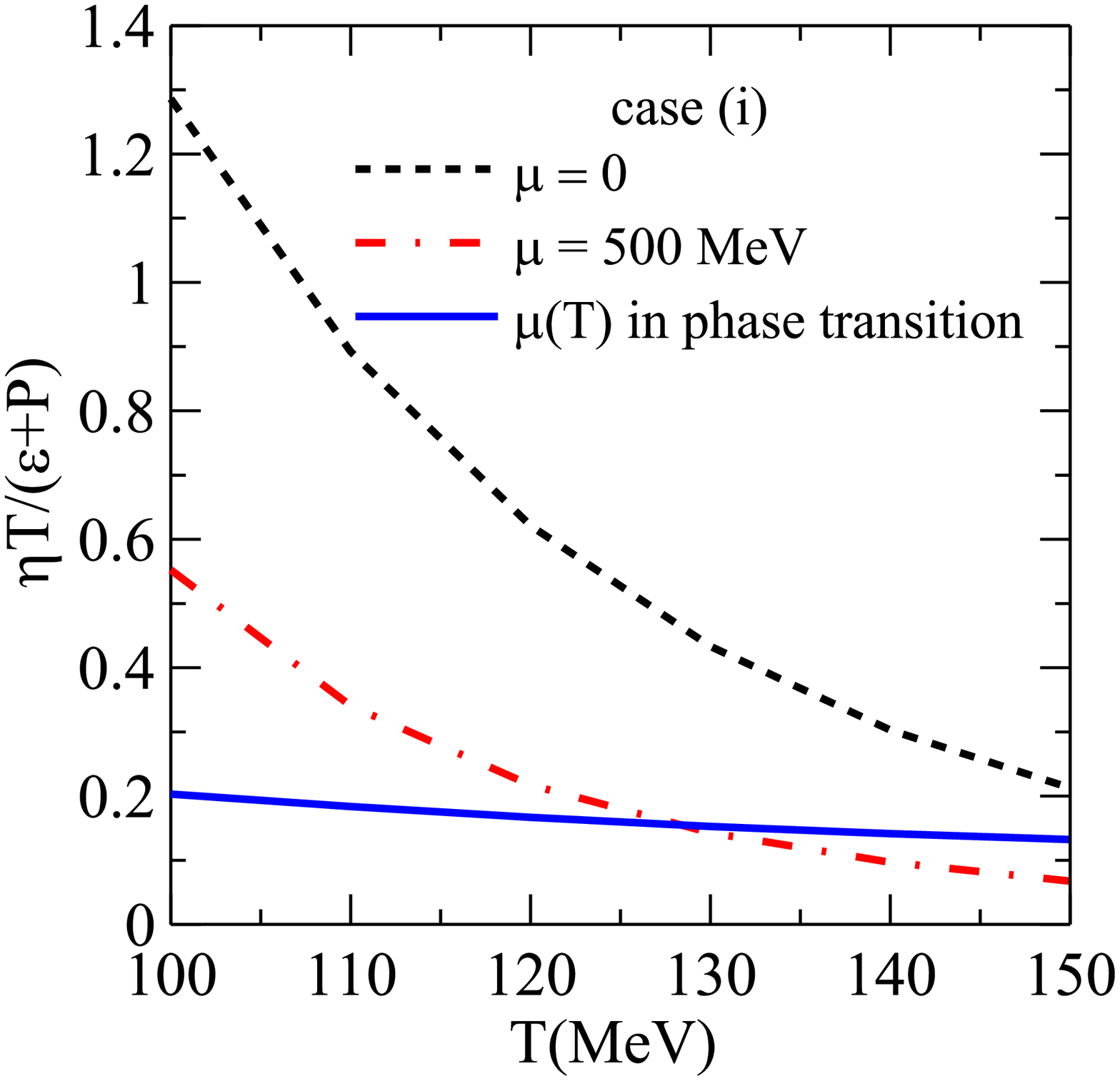} 
\end{minipage} 
\caption{The ratio $\protect\eta /s$ (left panel) and the ratio $\protect%
\eta /h$ (right panel) as a function of temperature (in MeV) using cross
sections (i) for several values of baryon chemical potential: $\protect\mu %
_{B}=0$ (black dotted line), $\protect\mu _{B}=500$ MeV (red dash-dotted
line), and $\protect\mu _{B}(T)$ in the chemical freeze-out transition. The
function $\protect\mu _{B}(T)$ is extracted from Ref.\ \protect\cite%
{Cleymans:2005xv}}
\label{fig1}
\end{figure}

In Fig.\ \ref{fig1} we show $\eta /s$ (left panel) and $\eta T/h$ (right
panel) as function of $T$ for the $3$ baryon chemical potential choices
described above and for the cross section of case (i). The black dotted line
corresponds to $\mu _{B}=0$, the red dash-dotted line corresponds to $\mu
_{B}=500$ MeV, and the blue solid line corresponds to $\mu _{B}(T)$ from
Ref. \cite{Cleymans:2005xv}.

One can see that a finite baryon chemical potential can have a very large
effect on $\eta /s$. The effect is particularly large at the chemical
freeze-out transition where $\eta /s$ is reduced by a factor $\sim 3$ at $%
100 $ MeV. As discussed in Ref.\ \cite{koch} and in the Introduction of this
paper, at nonzero baryon chemical potential $\eta /s$ is not the correct
quantity to estimate the fluidity of the system. In this case, the ratio $%
\eta T/h$ can provide a better estimate. We see the effect of $\mu _{B}$ is
amplified in this case, serving to reduce even more $\eta T/h$. For the case
of $\mu _{B}(T)$ defined using the chemical freeze-out transition, we see
that $\eta T/h$ is almost constant in temperature, which indicates that the
fluid behavior of a hadron resonance gas does not change much in the phase
transition region. This suggests that the systems created in the RHIC low
energy runs can exhibit substantial elliptic flow since $v_{2}(p_{T})$
should be dominated by the shear viscosity near the phase transition \cite%
{niemi_PRL}.

We remark that this dependence of $\eta /s$ (or $\eta T/h$) on $\mu _{B}$
originates mostly from the thermodynamic variables, e.g. $s$ and $h/T$. The
shear viscosity coefficient itself increases with $\mu _{B}$, even though this effect is small. The
reduction of $\eta /s$ and $\eta T/h$ with $\mu _{B}$ happens mostly because 
$s$ and $h/T$ increase by a significant amount when $\mu _{B}$ is present
and is of the order $\mu _{B}\sim 500$ MeV. In the Ref.~\cite{Hauer}, the
increase of $\eta /s$ with $\mu _{B}$ must have been due to the excluded
volume effect. We note, however, that in the equation of state usually
employed in heavy ion collision simulations (and that describes reasonably
well lattice QCD simulations) the excluded volume correction is not
included, for details see Ref.\ \cite{Huovinen}.

\begin{figure}[tbp]
\begin{minipage}{.45\linewidth}
\hspace{-1.5cm}
\includegraphics[width=1.1\textwidth]{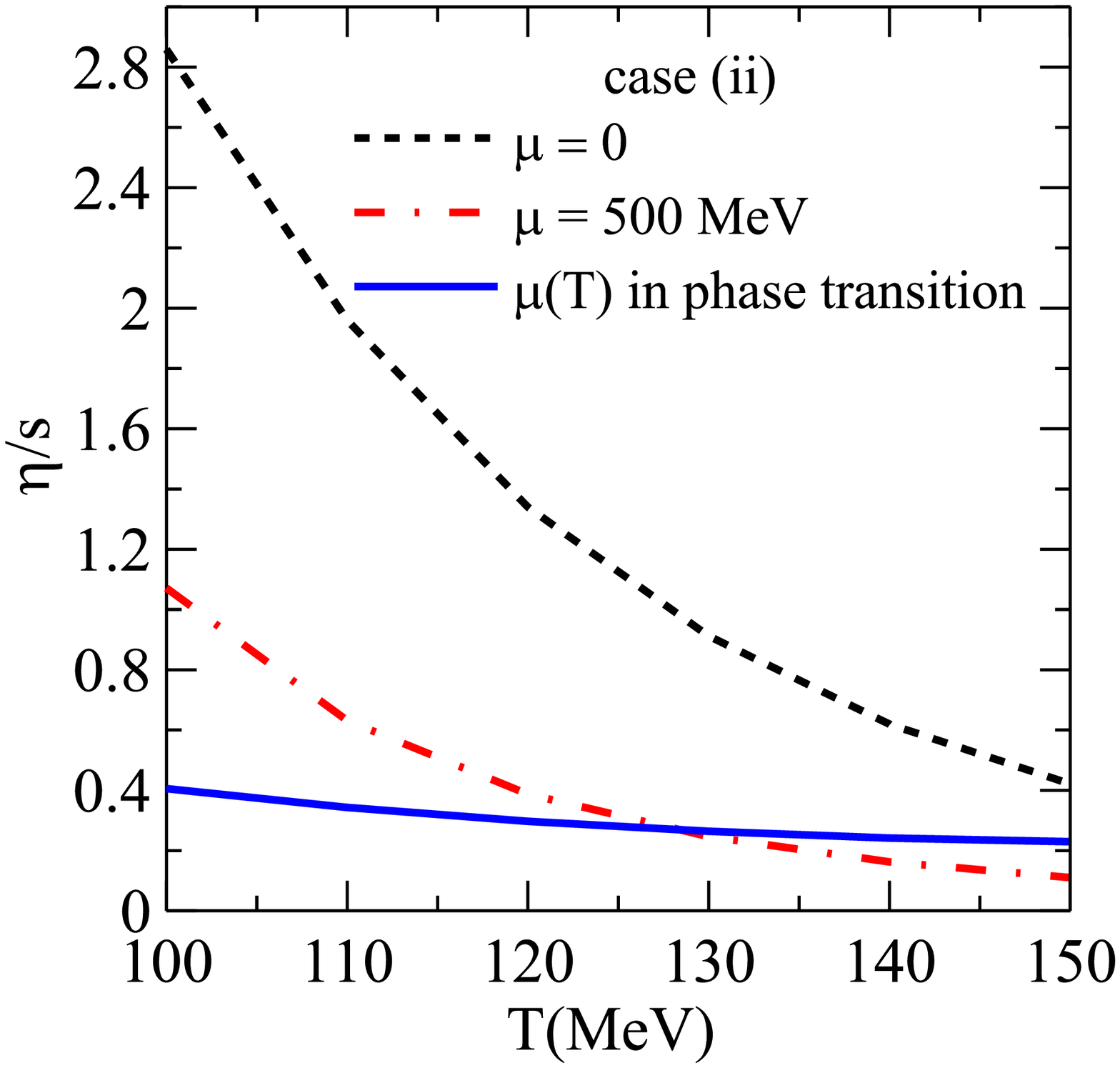} 
\end{minipage}  
\begin{minipage}{.45\linewidth}
\hspace{-1.5cm}
\includegraphics[width=1.1\textwidth]{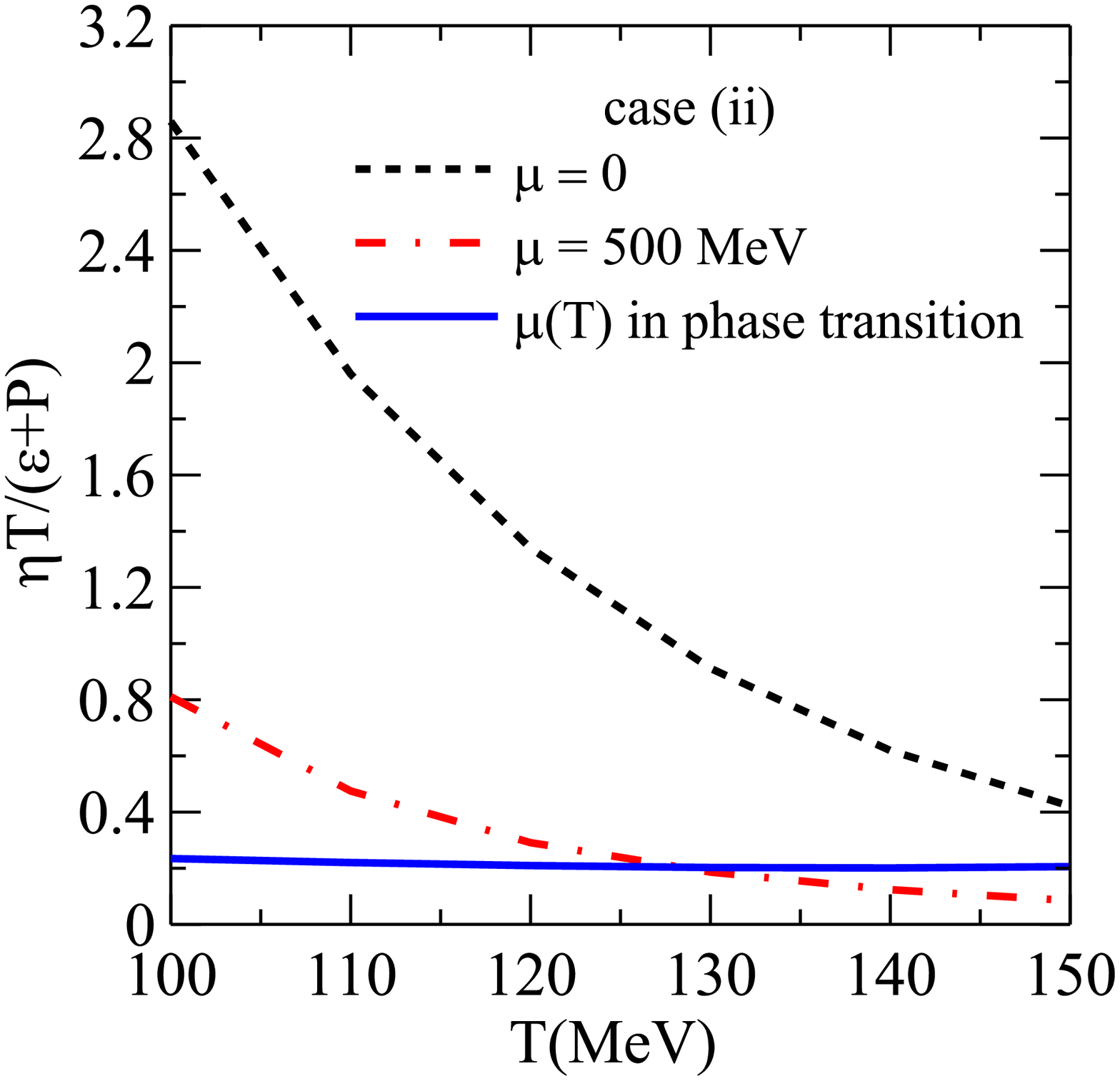} 
\end{minipage} 
\caption{The ratio $\protect\eta /s$ (left panel) and the ratio $\protect%
\eta /h$ (right panel) as a function of temperature (in MeV) using cross
sections (ii) for several values of baryon chemical potential: $\protect\mu %
_{B}=0$ (black dotted line), $\protect\mu _{B}=500$ MeV (red dash-dotted
line), and $\protect\mu _{B}(T)$ in the chemical freeze-out transition. The
function $\protect\mu _{B}(T)$ is extracted from Ref.\ \protect\cite%
{Cleymans:2005xv}}
\label{fig1.2}
\end{figure}

In Fig.\ \ref{fig1.2} we show $\eta /s$ (left panel) and $\eta T/h$ (right
panel) as function of $T$ for the $3$ baryon chemical potential choices
described above and for the cross section of case (ii). The results are
qualitatively similar to those found in case (i). The main difference is
that $\eta /s$ and $\eta T/h$ decrease faster with increasing temperature
for cross sections (ii) since the meson-meson cross section is almost a
factor two smaller than the baryon-baryon cross section and, at low
temperatures $\sim 100$ MeV, the system is dominated by mesons (mostly
pions). This can be seen by comparing Figs.\ \ref{fig1} and \ref{fig1.2}.
One should note that, in the hadron resonance gas model, the thermodynamic
quantities are independent of the cross sections and, consequently, the
entropy density, energy density, and thermodynamic pressure do not change
going from case (i) to case (ii).

\begin{figure}[tbp]
\begin{minipage}{.45\linewidth}
\hspace{-1.5cm}
\includegraphics[width=1.1\textwidth]{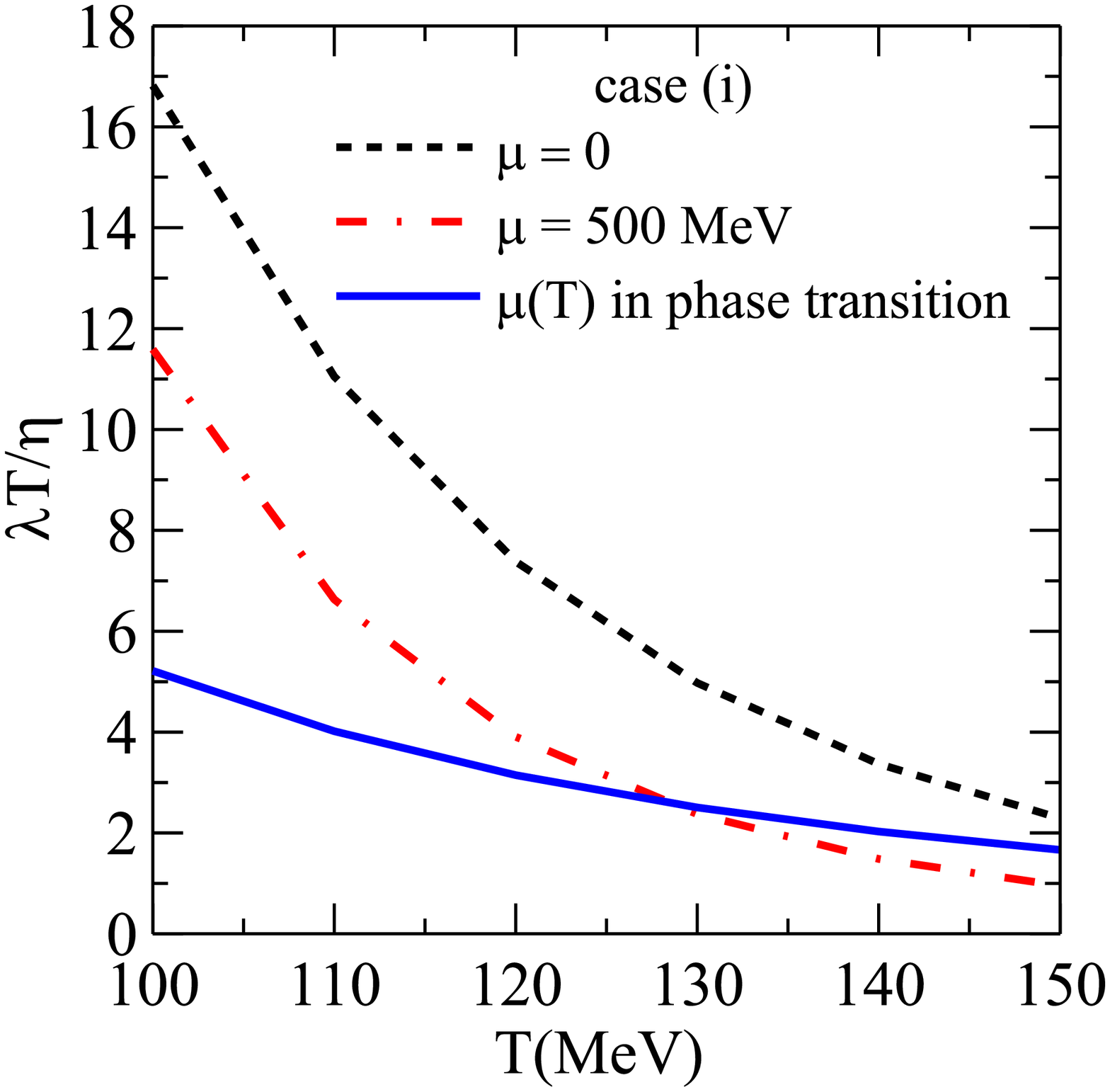} 
\end{minipage}  
\begin{minipage}{.45\linewidth}
\hspace{-1.5cm}
\includegraphics[width=1.1\textwidth]{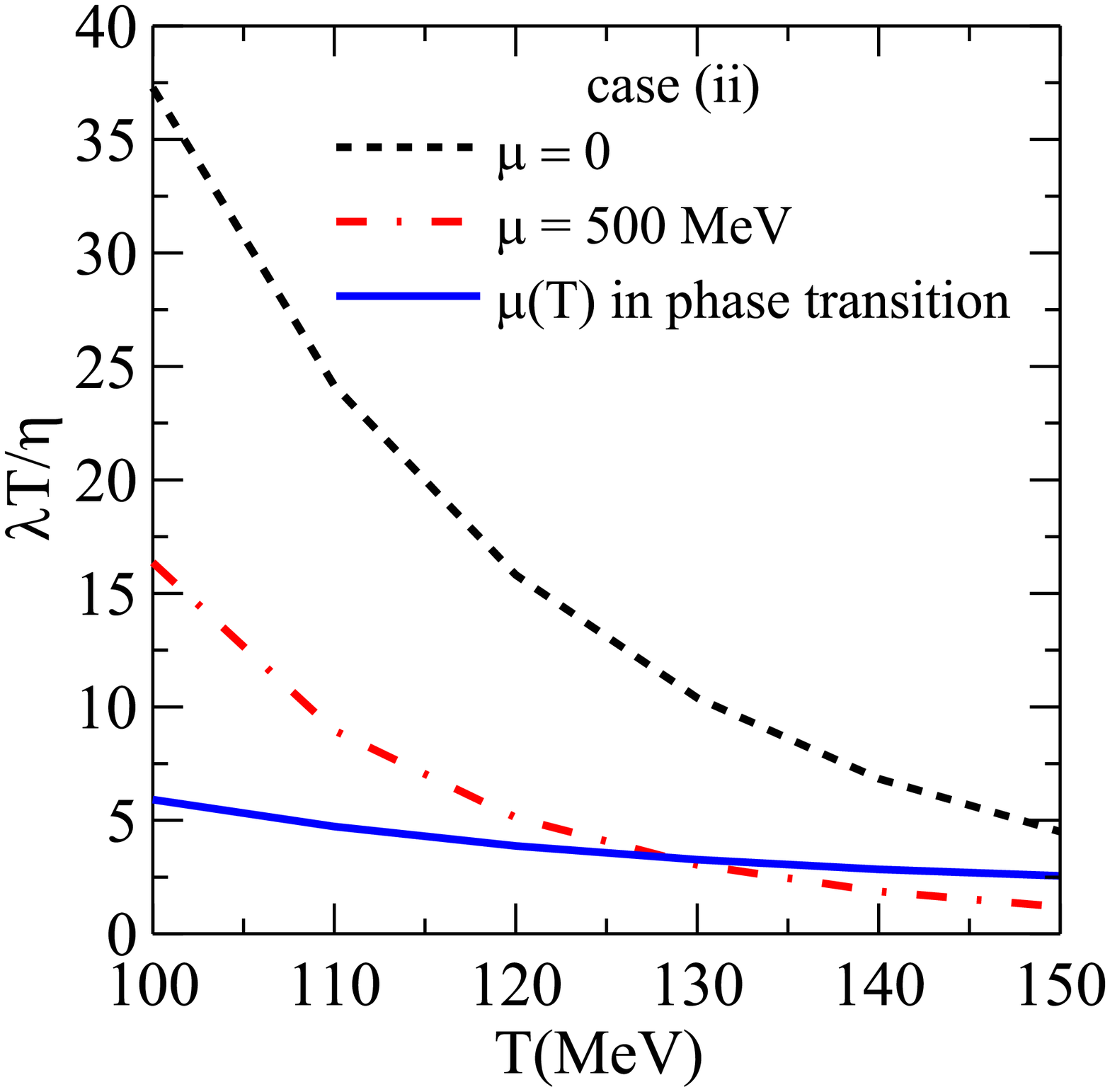} 
\end{minipage} 
\caption{Second order transport coefficient of the Burnett (gradient)
expansion, $\protect\lambda$, normalized by $\protect\eta/T $, as a function
of the temperature (in MeV) for several values of baryon chemical potential: 
$\protect\mu _{B}=0 $ (black dotted line), $\protect\mu _{B}=500$ MeV (red
dash-dotted line), and $\protect\mu _{B}(T)$ in the chemical freeze-out
transition. The function $\protect\mu _{B}(T)$ is extracted from Ref.\ 
\protect\cite{Cleymans:2005xv}. The left panel shows the results
corresponding to cross sections of case (i) while the right panel shows the
results corresponding to case (ii). }
\label{fig2}
\end{figure}

It is also useful to see what happens to the transport coefficients of terms
that are of second order in gradients. In Fig.\ \ref{fig2}, we show the
transport coefficient $\lambda $, that multiplies the next order (linear)
term in the Burnett equation (i.e., the gradient expansion), for cases (i)
(left panel) and (ii) (right panel). In this case, this is the coefficient
that multiplies the term $\dot{\sigma}^{\mu \nu }$. We plot the following
dimensionless combination $\lambda T/\eta $ in Fig.\ \ref{fig2} and one can
see that this dimensionless ratio is also reduced when baryon chemical
potential is present. From a qualitative perspective, the effect is very
similar to what happens to $\eta T/h$. The result is qualitatively the same
for both choices of cross sections.

This shows that the fluidity of a baryon rich hadron gas is in fact larger
than the equivalent system at vanishing baryon number. Not only $\eta
T/\left( \varepsilon +p\right) $ is considerably smaller, but the
dimensionless combination $\lambda T/\eta $, associated with a term of
second order in gradients of velocity, becomes considerably smaller when $%
\mu _{B}$ is nonzero.

\section{Conclusion}

In this paper we investigated the effects of a nonzero baryon chemical
potential on the transport properties of a hadron resonance gas. We found
that a hadron resonance gas with large baryon number density is closer to
the ideal fluid limit than the corresponding gas with zero baryon number. A
nonzero baryon chemical potential served not only to reduce the effect of
dissipative terms of first order in gradients but also of terms that are of
second order. This suggests that the system created at RHIC at lower
collision energies may display a fluid-like behavior with an effective
fluidity close to the one found at RHIC's highest energy collisions. This
may explain why the differential elliptic flow coefficient measured at lower
collisional energies at RHIC is close to the one measured at high energies.



G.~S.~Denicol, S.~Jeon, and C.~Gale acknowledge support by the Natural
Sciences and Engineering Research Council of Canada. J.~Noronha thanks
Funda\c c\~ao de Amparo \`a Pesquisa do Estado de S\~ao Paulo (FAPESP) and
Conselho Nacional de Desenvolvimento Cient\'ifico e Tecnol\'ogico (CNPq) for
support.

\end{document}